\documentclass[prl,
reprint,
 amsmath,amssymb,
 aps,
]{revtex4-2}

\usepackage{graphicx}
\usepackage{dcolumn}
\usepackage{bm}

\begin{document}


\title{Ground State Properties of the Doped Kitaev-Heisenberg Chain: Topological Superconducting and Mott Insulating Phases Driven by Magnetic Frustration}

\author{Cli\`o Efthimia Agrapidis}
\affiliation{Faculty of Physics, University of Warsaw, Pasteura 5, PL-02093 Warsaw, Poland}
\affiliation{Institute of Solid State Physics, TU Wien, 1040 Vienna, Austria}

\author{Satoshi Nishimoto}
\email{s.nishimoto@ifw-dresden.de}
\affiliation{Department of Physics, Technical University Dresden, 01069 Dresden, Germany}
\affiliation{Institute for Theoretical Solid State Physics, IFW Dresden, 01171 Dresden, Germany}

\date{\today}

\begin{abstract}
We study the hole-doped Kitaev-Heisenberg chain using the density-matrix renormalization group. In the Kitaev-only limit, the bond-directional exchange itself promotes pairing, favoring spin-singlet and spin-triplet superconducting tendencies for antiferromagnetic and ferromagnetic Kitaev couplings, respectively, together with finite-size Majorana edge correlations suggestive of topological superconductivity. In the full Kitaev-Heisenberg chain, cooperative $J$ and $K$ exchanges broadly stabilize superconductivity, while competition between $J$ and $K$ induces a strong filling dependence and enables superconductivity even when both $J$ and $K$ are weak. At quarter filling, this competition produces a Mott insulator with spontaneous hopping dimerization. These results identify magnetic frustration as a common mechanism underlying superconducting and interaction-driven insulating phases in doped Kitaev systems.
\end{abstract}

\maketitle


{\it Introduction.}--- 
The Kitaev model is a paradigmatic frustrated spin system that hosts a gapless $\mathbb{Z}_2$ spin liquid at the isotropic point $K_x = K_y = K_z$~\cite{Kitaev2006,Hermanns2018}. In candidate Kitaev materials, additional interactions---most notably Heisenberg exchange and off-diagonal $\Gamma$ terms---often stabilize magnetic order~\cite{Winter2017}, motivating ongoing efforts to access Kitaev-driven quantum phases by tuning microscopic parameters or by carrier doping~\cite{Takagi2019}. While the undoped Kitaev--Heisenberg (KH) model has been extensively studied~\cite{Trebst2022,Matsuda2025}, its doped counterpart remains far less understood.

Early (slave-boson) mean-field studies proposed superconducting (SC) states upon doping, including topological ones, but their detailed character remains unsettled~\cite{You2012,Hyart2012,Okamoto2013,Schmidt2018,Burnell2011,Mei2012,Liu2016}. By contrast, beyond-mean-field approaches and recent density-matrix renormalization group (DMRG) studies have revealed strong competing tendencies and a sensitive dependence on geometry, interaction signs, and doping regime in doped Kitaev systems~\cite{Peng2021,Laurell2024}. A comprehensive, unbiased characterization of the doped KH model is therefore still lacking.

In this Letter, we present a systematic DMRG study of the hole-doped KH chain. We map out its ground-state phase diagram and identify (i) SC regimes whose spin structure depends on the signs of $J$ and $K$, together with finite-size signatures suggestive of topological superconductivity over a broad parameter range, and (ii) an interaction-driven dimerized Mott insulator at quarter filling ($n=0.5$) induced by magnetic frustration. Given the established correspondence between one- and two-dimensional phase diagrams in the Mott-insulating KH model~\cite{Chaloupka2013,Agrapidis2018,Catuneanu2019,Agrapidis2019}, our results provide useful guidance for doped Kitaev materials and higher-dimensional settings.

\begin{figure}[bt]
	\centering
	\includegraphics[width=1.0\linewidth]{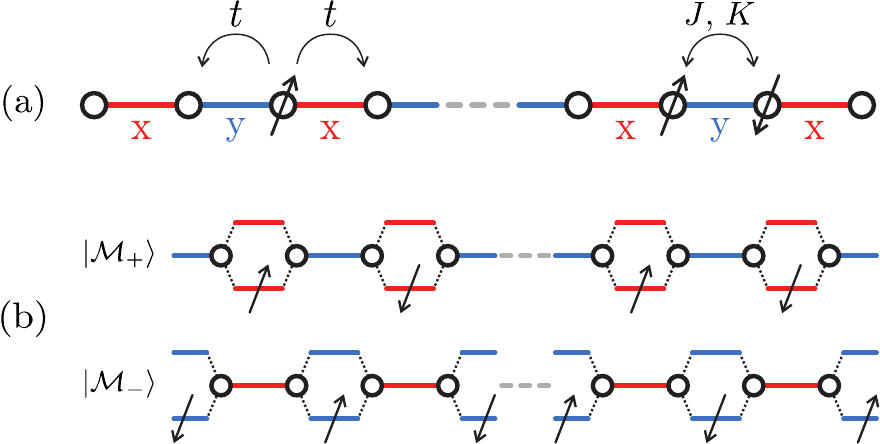}
	\caption{
		(a) Lattice structure of the KH chain.
		(b) Schematic picture of two translation-symmetry-breaking
		Mott insulating states, where each bonding orbital is occupied
		by one fermion.
	}
	\label{fig:lattice}
\end{figure}

{\it Model.}---
We study the hole-doped KH chain described by the projected $t$--$J$--$K$ Hamiltonian
\begin{align}
	\mathcal{H}=&-t\sum_{i=1}^{L}\sum_{\sigma}\left( f^{\dagger}_{i,\sigma} f_{i+1,\sigma}+{\rm H.c.}\right)
	+J\sum_{i=1}^{L}\mathbf{S}_i\cdot\mathbf{S}_{i+1}\nonumber\\
	&+K\sum_{i=1}^{L/2}\left(S^x_{2i-1}S^x_{2i}+S^y_{2i}S^y_{2i+1}\right),
	\label{eq:ham}
\end{align}
where $f^{\dagger}_{i,\sigma}$ is the electron creation operator with spin $\sigma$ at site $i$, acting in the projected Hilbert space with no double occupancy, and $\mathbf{S}_i=(S_i^x,S_i^y,S_i^z)$ is a spin-$1/2$ operator. The parameters $t$, $J$, and $K$ denote the hopping amplitude, Heisenberg exchange, and Kitaev exchange, respectively. As shown in Fig.~\ref{fig:lattice}(a), we refer to the bonds $(2i\!-\!1,2i)$ and $(2i,2i\!+\!1)$ as the $x$ and $y$ bonds, respectively.

{\it Methods and observables.}---
We investigate Hamiltonian~\eqref{eq:ham} using the DMRG method with open boundaries on chains of length up to $L=400$. We keep up to $m=12000$ states, with maximum truncation errors $\lesssim 10^{-4}$, and extrapolate observables to the $m\to\infty$ limit when necessary.

To map out the ground-state phase diagram, we extract the Tomonaga--Luttinger liquid (TLL) parameter $K_\rho$ from the long-wavelength limit of the static charge structure factor $N(q)=\frac{1}{L}\sum_{j,\ell} e^{iq(j-\ell)}\langle (n_j-n)(n_\ell-n)\rangle$ via $K_\rho=\pi \lim_{q\to 0} N(q)/|q|$~\cite{Ejima2005}. In a TLL, $K_\rho>1$ indicates dominant SC correlations. We also compute the pair-binding energy $\Delta_{\rm B}=E_0(N-2)+E_0(N)-2E_0(N-1)$, where $E_0(N)$ is the ground-state energy in the sector with $N$ electrons. A positive $\Delta_{\rm B}$ indicates effective attraction between doped holes.

To probe topological SC signatures, we evaluate the end-to-end Majorana correlator $G_{1L}= i\langle \lambda_{1,\sigma}\bar\lambda_{L,\sigma}\rangle$ with $\lambda_{j,\sigma}=f_{j,\sigma}+f^{\dagger}_{j,\sigma}$ and $\bar\lambda_{j,\sigma}=f_{j,\sigma}-f^{\dagger}_{j,\sigma}$~\cite{Miao2018}. For finite $L$, $G_{1L}$ can be nonzero due to edge-mode hybridization, although it vanishes in the thermodynamic limit. Its magnitude and parameter dependence nevertheless provide a useful finite-size diagnostic of regimes with pronounced edge-Majorana character.

To further characterize each phase, we compute the charge gap
$\Delta_{\rm c}=\{E_0(N+2)+E_0(N-2)-2E_0(N)\}/2$,
the connected density and spin correlations
$D(r)=\langle n_i n_{i+r}\rangle-\langle n_i\rangle\langle n_{i+r}\rangle$ and
$S^\alpha(r)=\langle S^\alpha_i S^\alpha_{i+r}\rangle-\langle S^\alpha_i\rangle\langle S^\alpha_{i+r}\rangle$,
and the pair--pair correlations $P_\gamma(r)=\langle\Delta^\dagger_{\gamma,i+r}\Delta_{\gamma,i}\rangle$.
Here $\Delta_{{\rm S},i}=(f_{i,\uparrow} f_{i+1,\downarrow}-f_{i,\downarrow} f_{i+1,\uparrow})/\sqrt{2}$ creates a spin-singlet pair, while
$\Delta_{{\rm T1},i}=(f_{i,\uparrow} f_{i+1,\downarrow}+f_{i,\downarrow} f_{i+1,\uparrow})/\sqrt{2}$ and
$\Delta_{{\rm T2},i}^{(\sigma)}=f_{i,\sigma} f_{i+1,\sigma}$ ($\sigma=\uparrow,\downarrow$) create spin-triplet pairs. Within TLL theory, $D(r)\propto r^{-K_\rho}$ and $P_\gamma(r)\propto r^{-1/K_\rho}$ at long distances.

\begin{figure}[tb]
	\centering
	\includegraphics[width=0.8\linewidth]{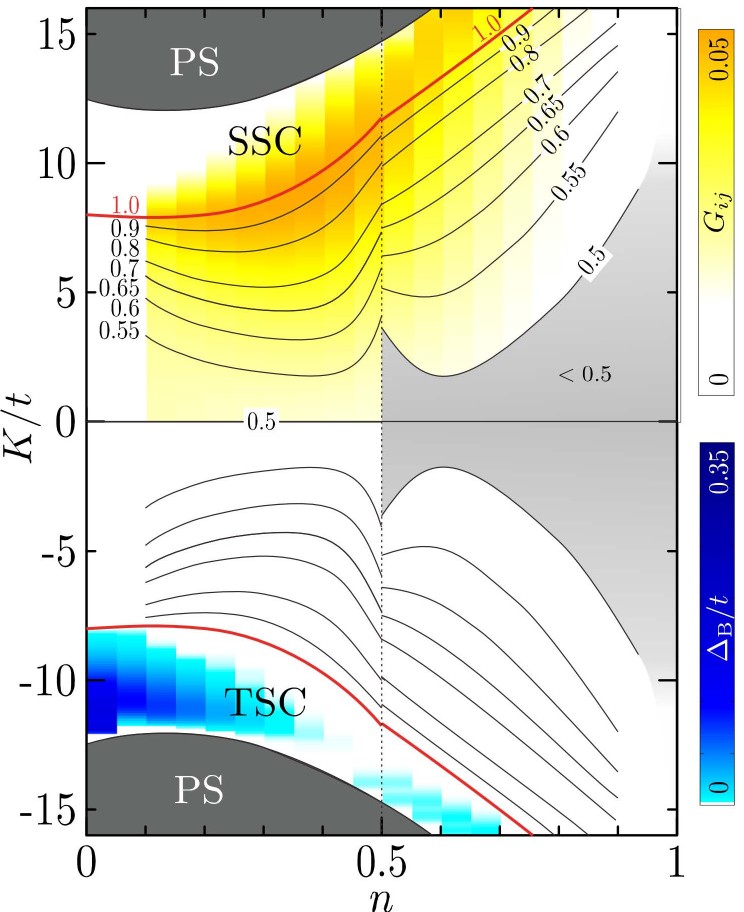}
	\caption{
		Phase diagram of the $t$--$K$ chain ($J=0$) in the $K/t$--$n$ plane. Contour lines indicate the TLL parameter $K_\rho$, while color maps show the end-to-end Majorana correlator $G_{1L}$ and the pair-binding energy $\Delta_{\rm B}/t$. Owing to the symmetry under $K\!\to\!-K$, only $G_{1L}$ for $K>0$ and $\Delta_{\rm B}$ for $K<0$ are shown (see text). The data for $G_{1L}$ are shown for an open chain with $L=40$, while those for $\Delta_{\rm B}$ are extrapolated to the thermodynamic limit.
	}
	\label{fig:J0PD}
\end{figure}

{\it Pairing induced by Kitaev interactions.}---
While Heisenberg exchange is known to promote superconductivity in doped Mott systems~\cite{Ogata1991,Moreno2011}, the pairing potential of the Kitaev interaction remains largely unexplored. To isolate its effect, we first focus on the $J=0$ limit and study the doped $t$--$K$ chain. The resulting ground-state phase diagram in the $K/t$--$n$ plane is shown in Fig.~\ref{fig:J0PD}, which exhibits an extended region with $K_\rho>1$, indicating dominant SC correlations within a TLL.

The physical origin of this pairing tendency can already be understood from a local two-particle viewpoint. Consider two carriers occupying neighboring sites, and denote the local two-spin wave function on a bond by $|\psi\rangle$. The Kitaev energy on a $\gamma$ bond ($\gamma=x,y$) is
$e_\gamma=\langle\psi|K S_i^\gamma S_j^\gamma|\psi\rangle$.
For antiferromagnetic (AFM) Kitaev exchange ($K>0$), both $e_x$ and $e_y$ are minimized by the singlet $|\psi^-\rangle=(|\uparrow\downarrow\rangle-|\downarrow\uparrow\rangle)/\sqrt{2}$, yielding $e_x=e_y=-K/4$. For ferromagnetic (FM) exchange ($K<0$), the minima are instead achieved by the $S^z=0$ triplet $|\psi^+\rangle=(|\uparrow\downarrow\rangle+|\downarrow\uparrow\rangle)/\sqrt{2}$, again giving $e_x=e_y=-|K|/4$. Thus, the SC regimes for $K>0$ and $K<0$ are naturally associated with spin-singlet SC (SSC) and spin-triplet SC (TSC) pairing, respectively. In this sense, the Kitaev interaction provides an effective short-range attraction between neighboring carriers and favors superconductivity at sufficiently large $|K|$, much like in the $t$--$J$ model. Upon further increasing $|K|$, however, the system eventually undergoes phase separation (PS), as is also well known in the $t$--$J$ model.

We note that the Hamiltonians with $K$ and $-K$ are unitarily related by a staggered $\pi$ spin rotation, which preserves the energy spectrum while mapping SSC to TSC. Although the present one-dimensional (1D) geometry contains only $x$ and $y$ bonds, it is instructive to note how this local argument extends when a $z$-bond is also present, as in higher-dimensional Kitaev lattices. For a $z$-bond one has $e_z=\langle\psi|K S_i^z S_j^z|\psi\rangle$. Since $\langle\psi^\pm|S_i^z S_j^z|\psi^\pm\rangle=-1/4$, the states $|\psi^\pm\rangle$ minimize $e_z$ only for $K>0$. For $K<0$, the $z$-bond energy is instead minimized by the equal-spin triplets $|\uparrow\uparrow\rangle$ and $|\downarrow\downarrow\rangle$, for which $\langle S_i^z S_j^z\rangle=+1/4$ and hence $e_z=-|K|/4$. This highlights the bond-selective anisotropy of Kitaev exchange beyond the present 1D setting.

To further characterize the SC regimes in Fig.~\ref{fig:J0PD}, we show the pair-binding energy $\Delta_{\rm B}/t$ in the thermodynamic limit and the end-to-end Majorana correlator $G_{1L}$ for an open chain with $L=40$. While $\Delta_{\rm B}$ is reduced around $n\simeq 0.5$, it remains finite over a broad portion of the pairing-dominated region. The correlator $G_{1L}$ serves as a finite-size diagnostic of edge-Majorana character: it can be nonzero for finite $L$ due to edge-mode hybridization and decreases toward zero in the thermodynamic limit. Nevertheless, the pronounced enhancement of $|G_{1L}|$ at larger fillings, away from the dilute limit, is suggestive of topological superconductivity.

\begin{figure}[tb]
	\centering
	\includegraphics[width=0.9\linewidth]{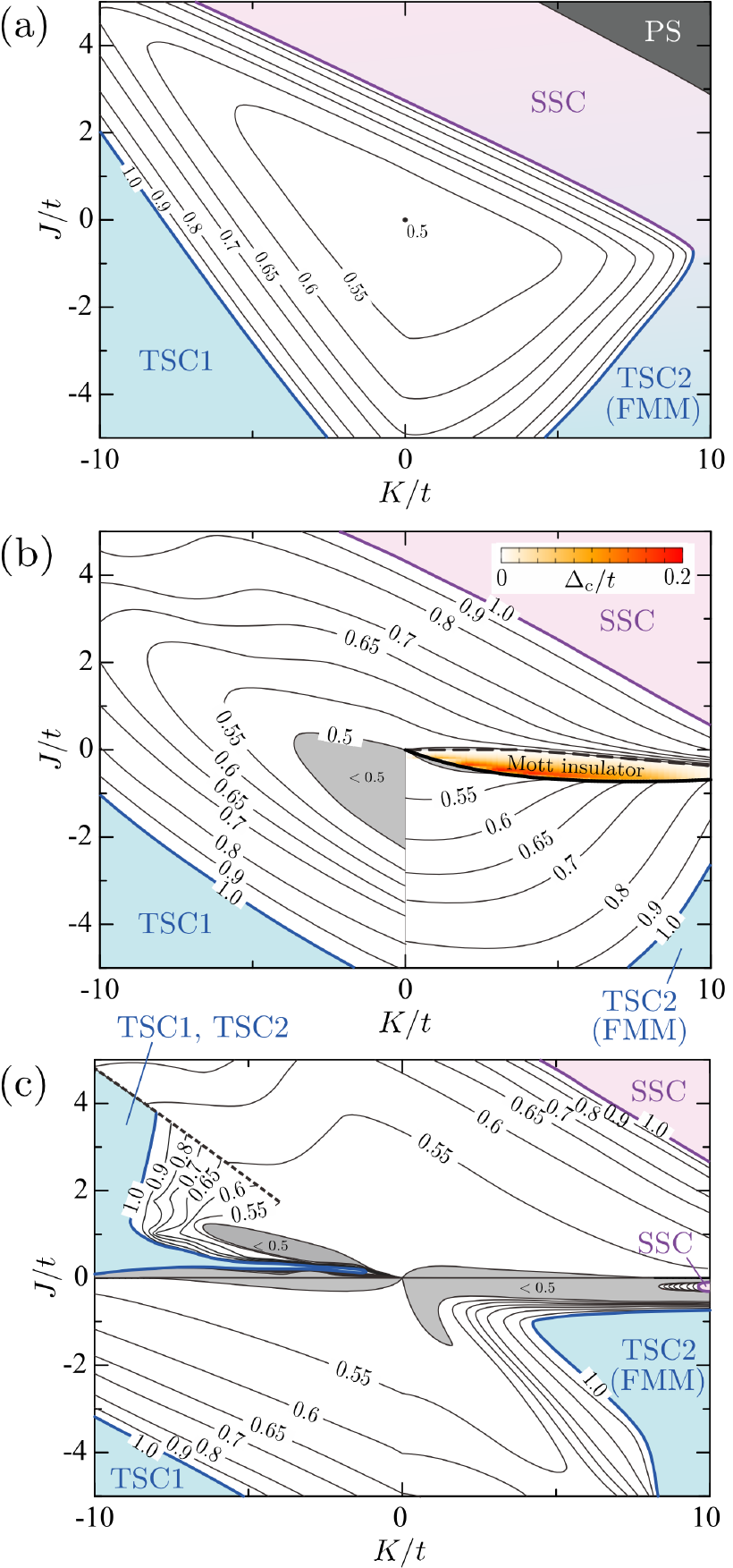}
	\caption{
		Phase diagram of the $t$--$J$--$K$ chain in the $K/t$--$J/t$ plane at (a) $n=0.1$, (b) $0.5$, and (c) $0.9$. A color map of the charge gap $\Delta_{\rm c}/t$ is shown for the Mott-insulating phase at $n=0.5$.
	}
	\label{fig:PD}
\end{figure}

{\it Ground-state phase diagram of the doped Kitaev--Heisenberg chain.}---
We now turn to the generic case with both Kitaev and Heisenberg exchanges present. Fig.~\ref{fig:PD} summarizes the ground-state phase diagram of the doped KH chain in the $K/t$--$J/t$ plane for representative fillings $n=0.1$, $0.5$, and $0.9$.

At all doping levels, a robust trend becomes apparent when $J$ and $K$ have the same sign. Here, the two exchanges cooperate to stabilize superconductivity once the overall exchange scale becomes sufficiently large, with relatively weak filling dependence of the phase boundary. For $J,K>0$ we find a SSC regime, whereas for $J,K<0$ a TSC regime is realized, as confirmed below by the pair--pair correlations.

By contrast, when $J$ and $K$ have opposite signs, the Kitaev- and Heisenberg-driven tendencies compete strongly, leading to pronounced filling-dependent behavior. Two salient features emerge in this regime. First, at $n=0.5$, a dimerized Mott-insulating state appears in the narrow region $K>0$, $J<0$, and $K+J>0$, where a finite charge gap $\Delta_{\rm c}$ opens [Fig.~\ref{fig:PD}(b)]. The transition on the small-$|J|$ side is consistent with a BKT-type transition, whereas that on the large-$|J|$ side appears to be first order. Second, for $n=0.9$, superconductivity develops already at relatively small $J/t$ over a broad range of both positive and negative $K/t$. In this case, the SC region ($K_\rho>1$) lies next to a strongly competing metallic regime with a substantially reduced TLL parameter, typically $K_\rho\lesssim 0.5$.

These results suggest that magnetic frustration plays a dual role in the doped KH chain: at commensurate filling it can drive an interaction-induced insulating state, while away from commensuration it enhances SC tendencies near the boundary of the strongly competing regime. More broadly, the pronounced suppression of $K_\rho$ indicates that the competition between Kitaev and Heisenberg exchanges strongly reshapes the low-energy charge response. Clarifying the microscopic origin of this effect deep inside the competing metallic regime remains an interesting open problem~\cite{Peng2021}.

In the following, we characterize the dimerized Mott-insulating state in detail and determine the spin structure of the SC regimes by comparing the relevant correlation functions.

\begin{figure}[tb]
	\centering
	\includegraphics[width=0.8\linewidth]{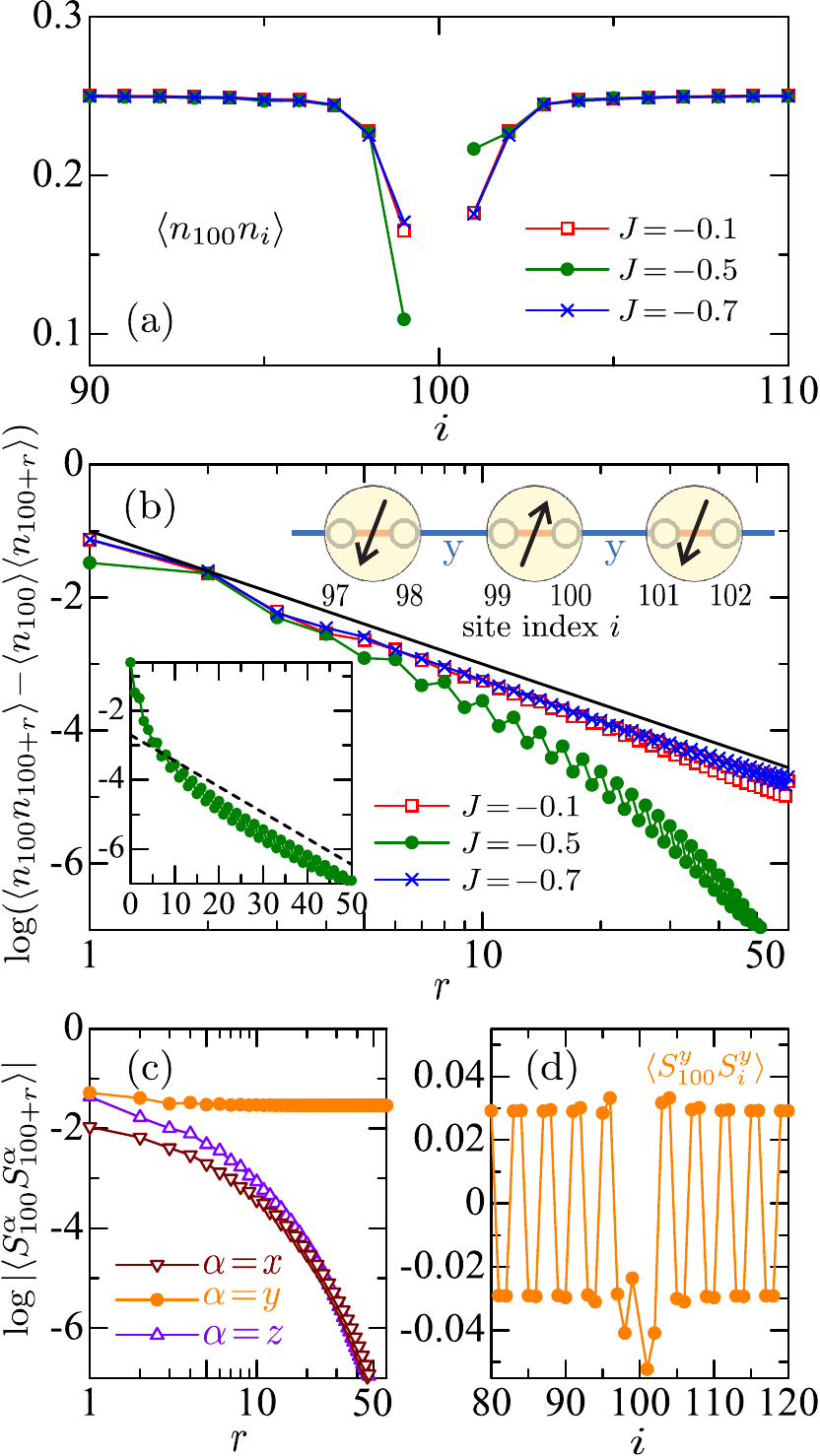}
	\caption{
		Results for an open chain of length $L=200$.
		(a) Density--density correlation $\langle n_{100} n_i\rangle$ at $K=5$.
		(b) Connected density--density correlation $D(r)=\langle n_{100} n_{100+r}\rangle-\langle n_{100}\rangle\langle n_{100+r}\rangle$ at $K=5$, shown on a log--log scale as a function of distance $r$. Insets: semi-log plot of $D(r)$ (left) and a schematic picture of the dimerized Mott state (right).
		(c) Spin--spin correlations $|\langle S^\alpha_{100} S^\alpha_{100+r}\rangle|$ for $\alpha=x,y,z$ at $K=5$ and $J=-0.5$.
		(d) Real-space profile of the $y$-component correlation $\langle S^y_{100} S^y_i\rangle$ at $K=5$ and $J=-0.5$.
}

	\label{fig:Mott}
\end{figure}

{\it Dimerized Mott state at quarter filling.}---
We now examine in more detail the dimerized Mott-insulating state realized at $n=0.5$ for $K>0$ and $J<0$. In this regime, the Kitaev and FM Heisenberg exchanges compete strongly, and the system relieves the resulting frustration by developing a spontaneous dimerization of the kinetic energy, i.e., an alternation of strong and weak bonds.

This dimerization is directly visible in real space from density--density correlations on an open chain. Figure~\ref{fig:Mott}(a) shows $\langle n_{100} n_i\rangle$ at $K=5$ for $J=-0.1$, $-0.5$, and $-0.7$ on a chain of length $L=200$. For $J=-0.1$ and $-0.7$, the correlations remain essentially left--right symmetric around $i=100$, indicating the absence of bond dimerization. By contrast, for $J=-0.5$ we observe a clear asymmetry between the nearest-neighbor correlations, $\langle n_{100} n_{99}\rangle < \langle n_{100} n_{101}\rangle$, consistent with an alternating strong--weak bond pattern [inset of Fig.~\ref{fig:Mott}(b)].

Once such a bond dimerization develops, the strong bonds form bonding and antibonding orbitals. At $n=0.5$, corresponding to one electron per two sites, the bonding orbitals are effectively half filled, so that a Mott gap opens within the dimerized background. This interpretation is consistent with both the finite charge gap $\Delta_{\rm c}$ and the behavior of the connected density correlation $D(r)$: as shown in Fig.~\ref{fig:Mott}(b), $D(r)$ decays algebraically for $J=-0.1$ and $-0.7$, as expected for a TLL, whereas for $J=-0.5$ it crosses over to exponential decay, signaling a charge-gapped state.

The dimerization pattern is twofold degenerate, corresponding predominantly to dimers on the $x$- or $y$-type bonds, as illustrated in Fig.~\ref{fig:lattice}(b). On open chains, the boundaries weakly pin one of these two patterns, producing the observed real-space texture. In addition, within this dimerized Mott phase we find genuine long-range order in the $y$-spin correlations, $\langle S_i^y S_j^y\rangle$ [Figs.~\ref{fig:Mott}(c) and \ref{fig:Mott}(d)], consistent with an effective low-energy description in terms of an Ising-like spin-$1/2$ chain dominated by the $(K+J)S_i^y S_j^y$ channel.

\begin{figure}[tb]
	\centering
	\includegraphics[width=0.9\linewidth]{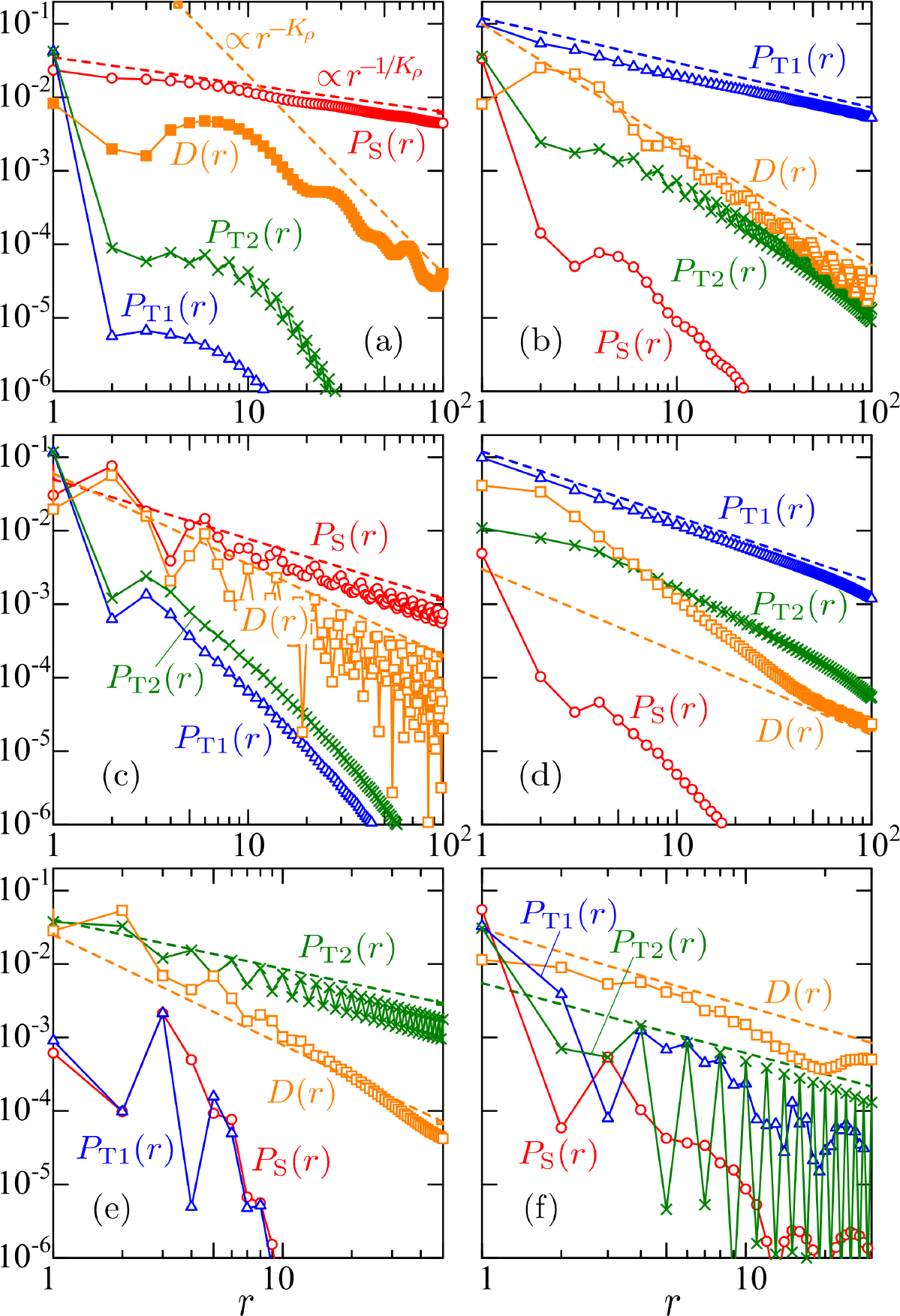}
\caption{
	Pair--pair correlation functions $P_\gamma(r)$ ($\gamma={\rm S, T1, T2}$) and density--density correlation functions $D(r)$, plotted on log--log scales for representative parameter sets in the phase diagram:
	(a) $K=10.0$, $J=0.0$, $n=0.1$ (SSC, $K_\rho=2.70$);
	(b) $K=-11.0$, $J=0.0$, $n=0.3$ (TSC1, $K_\rho=1.64$);
	(c) $K=5.0$, $J=3.0$, $n=0.5$ (SSC, $K_\rho=1.23$);
	(d) $K=-5.0$, $J=-4.0$, $n=0.5$ (TSC1, $K_\rho=1.13$);
	(e) $K=10.0$, $J=-5.0$, $n=0.5$ (TSC2, $K_\rho=1.50$); and
	(f) $K=-10.0$, $J=2.0$, $n=0.9$ (TSC1/TSC2, $K_\rho=1.05$).
	Panels (a)--(d) are obtained for open chains with $L=400$, while panels (e) and (f) are for $L=200$.
}
	\label{fig:corr}
\end{figure}

{\it Pairing symmetry from pair--pair correlations.}---
To identify the dominant SC correlations and their spin structure, we analyze the distance dependence of the pair--pair correlations $P_\gamma(r)$ for the singlet channel $\gamma={\rm S}$ and the triplet channels $\gamma={\rm T1,T2}$ defined in {\it Methods and observables}. Within a TLL, the dominant tendency is determined by the slowest-decaying correlation function at long distances. Comparing $P_\gamma(r)$ with the density--density correlation $D(r)$ therefore provides a direct way to identify the leading pairing channel, consistent with the criterion based on $K_\rho$.

Fig.~\ref{fig:corr} compares $P_{\rm S}(r)$, $P_{\rm T1}(r)$, $P_{\rm T2}(r)$, and $D(r)$ for representative points in the SC regions of the phase diagram, using open chains with $L=400$ [Figs.~\ref{fig:corr}(a)--(d)] and $L=200$ [Figs.~\ref{fig:corr}(e) and \ref{fig:corr}(f)]. For comparison, we also show the asymptotic TLL forms $r^{-1/K_\rho}$ and $r^{-K_\rho}$ using the values of $K_\rho$ extracted from $N(q)$.

In the SC regimes stabilized for $J,K>0$, the singlet correlation $P_{\rm S}(r)$ decays most slowly and dominates at long distances, identifying a SSC tendency. By contrast, for $J,K<0$, the leading correlations are triplet, indicating a TSC tendency. In the latter case, the Kitaev exchange further selects the ${\rm T1}$ channel over ${\rm T2}$. These results also support the energetic classification obtained in the $J=0$ limit.

More specifically, in the SC phase found for $K>0$ and $J<0$ at $n\le 0.5$, $P_{\rm T2}(r)$ is dominant, while $P_{\rm T1}(r)$ decays much more rapidly. Interestingly, this regime is accompanied by a partially spin-polarized FM metal (FMM), with a magnetization of several tens of percent. By contrast, in the SC regime found for $K<0$ and $J>0$ at $n=0.9$, $P_{\rm T1}(r)$ and $P_{\rm T2}(r)$ are comparably enhanced and remain nearly degenerate over the accessible distance range. Although $D(r)$ is larger at short distances for this representative point, this mainly reflects a nonuniversal prefactor associated with local charge fluctuations; the slower asymptotic decay of $P_\gamma(r)$ identifies the dominant pairing tendency in the TLL regime.

Overall, these results establish a correlation-based classification of the SSC and TSC regions in Fig.~\ref{fig:PD}.

{\it Summary.}---
We have studied the hole-doped KH chain using DMRG and determined its ground-state phase diagram. In the Kitaev-only limit, we found that the Kitaev exchange itself promotes pairing, favoring spin-singlet and spin-triplet SC tendencies for AFM and FM Kitaev couplings, respectively, together with finite-size Majorana edge correlations suggestive of topological superconductivity. In the full Kitaev--Heisenberg chain, cooperative $J$ and $K$ exchanges broadly stabilize superconductivity when the overall exchange scale becomes sufficiently large, while competition between them induces a strong filling dependence and allows superconductivity to emerge even when both $J$ and $K$ couplings are weak. At quarter filling, we identified a Mott-insulating phase with spontaneous hopping dimerization. Our results highlight magnetic frustration as a common mechanism underlying superconducting and interaction-driven insulating phases in doped Kitaev systems.

\begin{acknowledgments}
We thank Ulrike Nitzsche for technical support. This work was supported by Narodowe Centrum Nauki (NCN, Poland) under Project No. 2021/40/C/ST3/00177. This research was carried out with the support of the Interdisciplinary  Center for Mathematical and Computational Modeling at the University of Warsaw (ICM UW) under Grants No. G81-4 and No. G93-1613. This project is funded by the German Research Foundation (DFG) via the projects A05 of the Collaborative Research Center SFB 1143 (Project No. 247310070).
\end{acknowledgments}

\bibliography{doped1DKH}

\end{document}